\documentclass{nature}
\pdfoutput=1
\usepackage{amsmath}
\usepackage{esint}
\usepackage{amssymb}
\usepackage{physics}
\usepackage[pdftex]{graphicx} 
\usepackage{caption}
\usepackage{siunitx}
\usepackage[utf8]{inputenc}
\usepackage{textcomp}
\makeatletter

\let\saved@includegraphics\includegraphics
\AtBeginDocument{\let\includegraphics\saved@includegraphics}
\renewenvironment*{figure}{\@float{figure}}{\end@float}

\def\blfootnote{\xdef\@thefnmark{}\@footnotetext}
\makeatother

\title{Temporally Resolved Intensity Contouring (TRIC) for characterization of the absolute spatio-temporal intensity distribution of a relativistic, femtosecond laser pulse}

\author{Daniel Haffa$^{1,*}$, Jianhui Bin$^{1,3,*}$, Martin Speicher$^{1,*}$, Klaus Allinger$^1$, Jens Hartmann$^1$, Christian Kreuzer$^1$, Enrico Ridente$^{1,2}$,  Tobias M. Ostermayr$^{1,3}$ and J\"org Schreiber$^1$}

\begin{document}

\maketitle

\begin{affiliations}
 \item Lehrstuhl f\"ur Medizinphysik, Fakult\"at f\"ur Physik, Ludwig-Maximillians-Universit\"at M\"unchen, 85748 Garching b. M\"unchen, Germany
 \item Max-Planck-Institut f\"ur Quantenoptik, 85748 Garching b. M\"unchen, Germany
 \item Accelerator Technology and Applied Physics Division, Lawrence Berkeley National Laboratory, CA 94720, USA
\end{affiliations}
\blfootnote{*Corresponding authors}
\begin{abstract}
	
Today's high-power laser systems are capable of reaching photon intensities up to $10^{22}$ \si[mode=text]{W.cm^{-2}}, generating plasmas when interacting with material. The high intensity and ultrashort laser pulse duration (fs) make direct observation of plasma dynamics a challenging task. In the field of laser-plasma physics and especially for the acceleration of ions, the spatio-temporal intensity distribution is one of the most critical aspects. We describe a novel method based on a single-shot (i.e. single laser pulse) chirped probing scheme, taking nine sequential frames at framerates up to THz. This technique, to which we refer as temporally resolved intensity contouring (TRIC) enables single-shot measurement of laser-plasma dynamics. Using TRIC, we demonstrate the reconstruction of the complete spatio-temporal intensity distribution of a high-power laser pulse in the focal plane at full pulse energy with sub picosecond resolution.

\end{abstract}

Since the advent of chirped pulse amplification\cite{strickland_compression_1985, maine_generation_1988} high-power laser systems have evolved and now enable novel particle acceleration techniques. Multi-GeV electrons\cite{esarey_physics_2009}, x-rays \cite{khrennikov_tunable_2015}, neutrons\cite{roth_bright_2013} and energetic ions\cite{kim_radiation_2016,wagner_maximum_2016}  emerge from intense laser-plasma interactions with targets. A milestone for acceleration of multi MeV ions has been demonstrated 2000\cite{snavely_intense_2000}. Short acceleration lengths (MeV/µm), small source sizes (µm), high particle flux\cite{daido_review_2012, macchi_ion_2013}, poly-energetic energy distributions and broad angular divergence\cite{schreiber_invited_2016} mark the key differences compared to the output from conventional accelerators. Substantial improvements of laser and target technology have enabled higher repetition rates also for solid targets. With the new generation of petawatt class laser systems\cite{papadopoulos_apollon_2016,_cala_} proton energies up to 100 MeV\cite{higginson_near-100_2018} in a single-shot and repetition rates up to 1 Hz\cite{gao_automated_2017} at lower kinetic energies are now available, fueling enthusiasm on the way to developing laser-driven systems\cite{schreiber_invited_2016}.

Two of the most influential aspects relevant to ion acceleration are the temporally resolved intensity contrast\cite{mckenna_high-intensity_2006} and the transverse intensity distribution in the focal plane. Due to sub-picosecond duration and the high peak intensity a direct measurement of such remains challenging. 
In fact the spatial and temporal intensity distributions are often determined independently, for separate shots and with attenuated beams. The temporal intensity contrast is typically measured relatively (in relation to the peak intensity) with auto correlation methods\cite{blount_recovery_1969, luan_high_1993} which integrate over the spatial distribution. Newest approaches aiming for single-shot measurements\cite{wang_single-shot_2015, oksenhendler_high_2017} seem very promising for monitoring the temporal shape of the intensity during the interaction. The spatial intensity distribution is usually measured with time-integrating beam profilers using an attenuated laser pulse and the anticipated high intensities require mapping with high dynamic range (at least five orders of magnitude) in order to avoid an overestimation of the peak intensity at focus\cite{hartmann_spatial_2018}.

A combined spatio-temporal measurement of an attenuated high-power laser focus has recently been demonstrated for the first time\cite{pariente_spacetime_2016}. Although several techniques allow the analysis of the spatial and temporal intensity distribution of a high-power laser focus, it has not been directly accessible during a laser-plasma interaction. Here we describe a technique that measures the spatio-temporal evolution of a laser-induced plasma on target for a single shot. This not only yields information about the evolution of plasma formation, but can be further interpreted to retrieve the absolute spatio-temporal intensity distribution (STID) of a high-power laser in the focal plane at full intensity. Within a single shot we measure temporally resolved intensity contours with sub-picosecond temporal resolution and about 25 µm spatial resolution. With each additional measurement with modified laser intensity (in our example five pulses) we add $ \sim $one order of magnitude to the covered dynamic range of our measurement. Because there are no direct comparisons with our novel technique, we extract the transverse spatial and the temporal profiles of the three-dimensional intensity distribution function, and compare them to the standard techniques, i.e. a spatial high dynamic range (HDR) focus image (integrated over the full laser pulse duration) and a temporal contrast measurement taken with a third-order auto-correlator.

\subsection{Experimental Setups}
\begin{figure}
	\includegraphics[width=\textwidth]{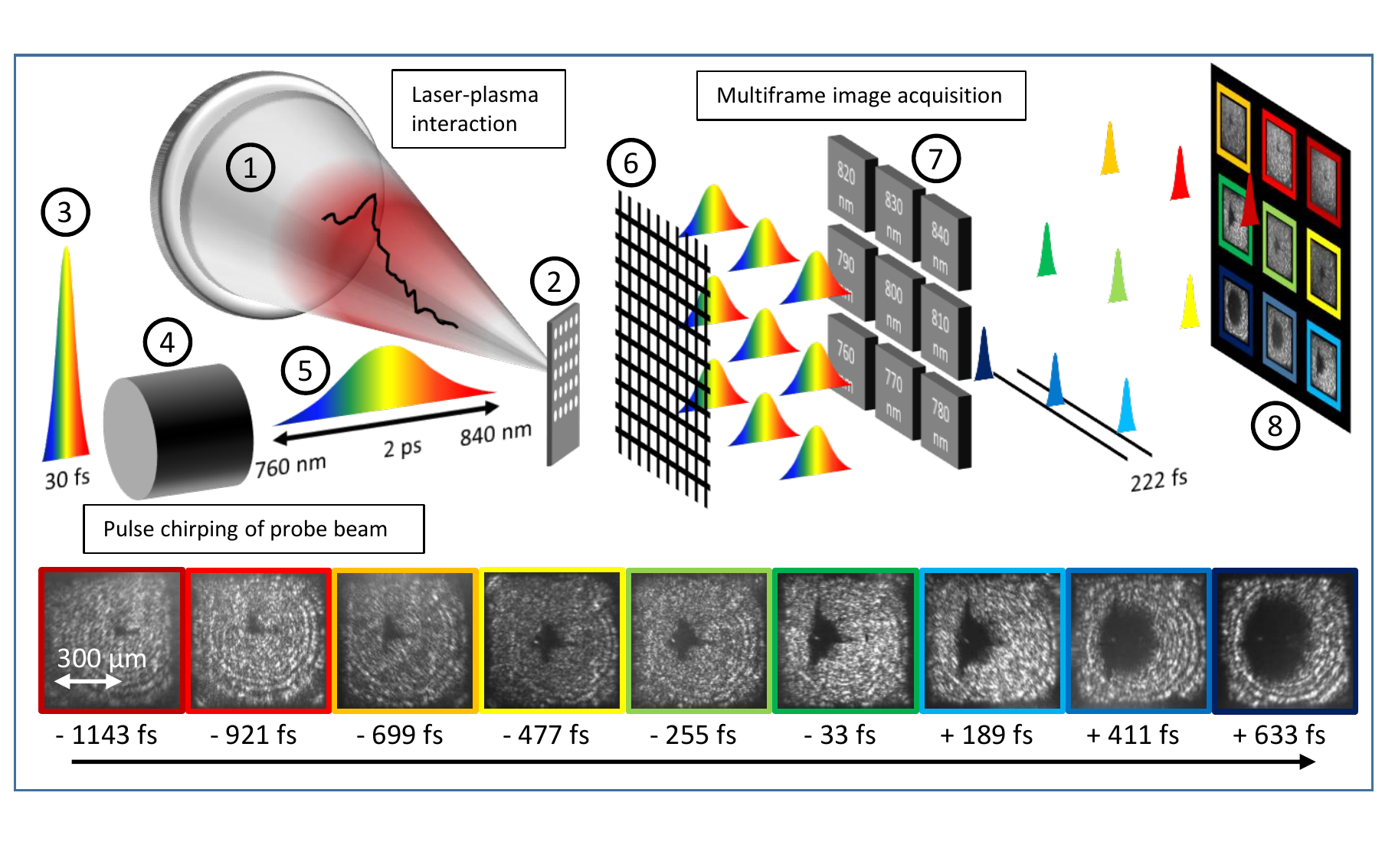}
	
	\caption{\textbf{Experimental Setup for TRIC}. A Ti:Sa laser pulse (\textbf{1}) is focused onto a 200 nm thick Formvar target (\textbf{2}) at $ 45^\circ $ incidence. A small part of the short laser pulse is coupled out earlier (\textbf{3}) and sent through a glass rod (\textbf{4}). The emerging chirped pulse (\textbf{5}) passes the target perpendicular to the main pulse. In the imaging path, the probe beam is multiplied using a low dispersion transmission grating (\textbf{6}). A small frequency range is cut out of each of the replicas by narrow bandpass filters (\textbf{7}) before being recorded with a camera (\textbf{8}). The bottom row shows a sample picture series recorded in a single-shot.}
	\label{Fig_ExperimentalDesign}

\end{figure}

The experimental setup resembles a typical pump-probe configuration and is shown in Fig. \ref{Fig_ExperimentalDesign}. The Ti:Sapphire system provides 374 mJ on target within 30 fs with a spectral range of 760 nm to 840 nm. The laser is focused onto a 200 nm thick Formvar foil target\cite{seuferling_efficient_2017} with a diameter of 2 mm which is positioned for irradiation at 45 degree incidence. A small fraction of the 30$ fs $ laser beam is coupled out in advance to provide the probe beam. After traversing a glass rod, the resulting chirped probe pulse of 2 ps duration is overlapped perpendicularly with the main laser pulse at the target. The transmitted probe light is collected with a lens, guided from the vacuum chamber and imaged with another lens onto a camera (Prosilica GT 4907, Allied Vision). A 2D grating (Collischon, 15 µm lattice constant) is positioned between the lens and the CCD chip, such that nine spatially separated replicated images are accommodated on the camera chip. Each image still contains the complete spectrum. By adding nine different narrow band-pass filters (band-width of 10 nm) in front of the camera chip at the positions of the replicas they will be spectrally filtered and thus correlated temporal information is imprinted\cite{gabolde_single-frame_2008,nakagawa_sequentially_2014} onto each image. The time delay between subsequent images is defined by the set spectral chirp and in our case set to 222 fs (see methods). We thus measure nine sequenced images with a frame rate of 5 THz, revealing the plasma evolution in response to a high-power laser pulse interacting with a solid density target. It is worth mentioning, that this implementation can considerably simplify single-shot probe schemes that have been realized in the past\cite{siders_efficient_1998, kaluza_time-sequence_2008, green_single_2014}. An exemplary raw image of such measurement is displayed in Fig.\ref{Fig_ExperimentalDesign}.

The image appears bright in areas where the target remains transparent and dark in areas, where the target became opaque. The cause for this binary image information is the change in the optical transmission of the evolving plasma. When the free electron density becomes larger than the critical density $n_e > n_c$, the plasma turns reflective for the incoming probe pulse. The horizontal dimension recorded in the image contains convoluted information since the laser and the orthogonal probe pulse both hit the target under an angle of 45 degree. The probe image is thus a projection. At the same time, the orientation of the target with respect to the drive-beam implies that non-central parts of the target interact with the laser in out-of-focus planes causing a complex two-fold convolution of time and space. A spatial difference in horizontal dimension of 60 µm in the image, corresponds to a time difference of 200 fs of both the laser and probe pulse hitting the target. For simplicity we therefore only consider the vertical dimension, where those effects do not play a role.

\subsection{Relation of laser intensity and plasma shape}
\begin{figure}
	\includegraphics[width=\textwidth]{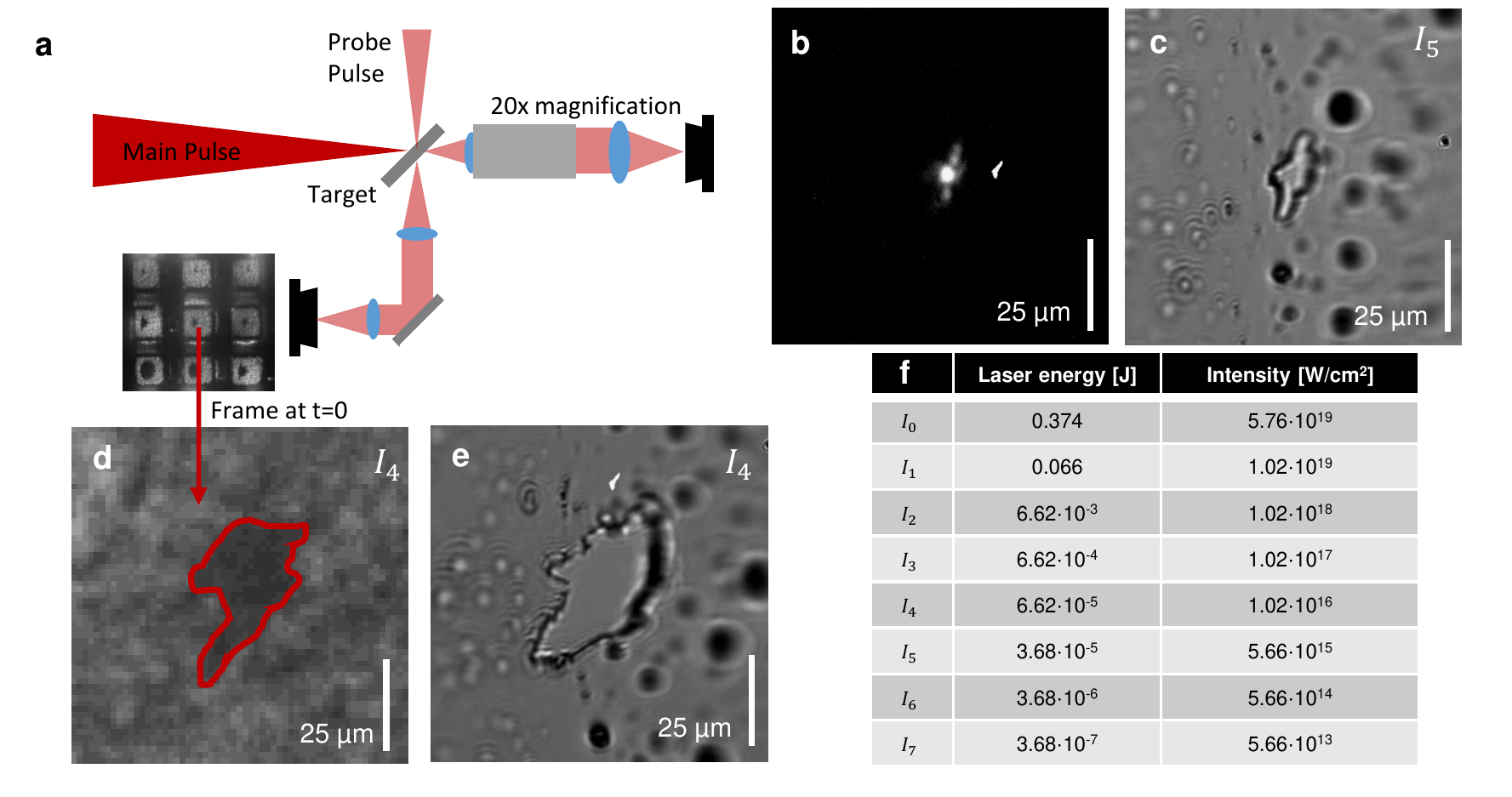}
	\caption{\textbf{Nexus of damage threshold and ionization threshold}. This figure shows separately obtained images of the laser focus, the probe image and the damage at the target after the shot. The configuration can be seen in \textbf{a}, a 20 times magnifying microsope views the target under an angle of $ 45^\circ $ and also measures an attenuated pulse in the focal plane \textbf{b}. The damage at the target with intensity $ I_5 $ is shown in \textbf{c}. A shot with intensity $ I_4 $ compares the damage of the target \textbf{e} to the probe image \textbf{d}. The table \textbf{f} lists the laser energies on target and corresponding intensities used during the experiment.}
	\label{Fig_Focus_DamageThr}
\end{figure}
The measured average plasma growth rate of the single-shot image-sequence in Fig. 1 is evaluated to be about $ 30\% $ of the speed of light. But as we will show, this is not due to kinetic plasma expansion. Since the majority of the plasma growth (in vertical dimension) took place prior to arrival of the main pulse (time t=0) and its extent remained unchanged for several 10s of picoseconds afterwards, we examined the source of the growing over-dense area in more detail. 

We varied the laser intensity and compared the spatial intensity distribution of the laser focus with the plasma size during the interaction and the damage (hole) at the target after the shot. The laser energy was varied by adding neutral density filters in the beam-path of the laser before compression and thus attenuating both the main pulse and the probe pulse. The corresponding intensities are shown in Fig. 2f. The damage in the target after the shot with intensity $ I_5 $ is shown in Fig 2c. Since the angle of the target was 45 degree with respect to the laser axis and thus also to the microscope (Fig 2a), only the central vertical axis appears sharp. Comparison to the focal image of the attenuated beam (Fig. 2b) indicates great resemblance. The picture was taken with the same microscope. Comparison of the laser intensity profile (Fig. 2b) and the damage spot (Fig. 2c) indicates the shape of the damage to be directly correlated to the shape of the laser focus. The laser focus was distorted to make this effect more clear. In Fig. 2d we see the measured plasma shape with the probing technique after t=0 (largest contour) with an intensity of $I_{4}$ of the main pulse. The hole in the target foil after the shot is shown in Fig. 2e and again the shapes in Fig. 2d and Fig. 2e show excellent resemblance. This indicates that the ionization threshold of the Formvar target can be related to its damage threshold\cite{soong_experimental_2011}. We hence assume that the measured plasma contour is the isosurface of the laser focus intensity reaching the damage threshold and thus establishing abruptly an overdense plasma. One source of temporal measurement uncertainty is the time it takes the plasma to evolve from being underdense to overdense. The damage threshold $I_{th}= 6 \pm 5 \cdot 10^{13}$ \si[mode=text]{W.cm^{-2}} (see methods) of the used material was evaluated by reducing the laser energy and thus the intensity until no damage (hole) was measurable at the target after the shot (see methods).

\section*{Results}

\subsection{Spatio-Temporal-Intensity-Distribution.}

The measured contour lines originate when the STID reaches the damage-threshold. The intensity at the edge contour is thus
\begin{equation} \label{eq_contour1}
I(x,y,t) = I_{th}.
\end{equation}

The intensity of the measured contour in the probe image is known but its inner intensity remains inaccessible due to the binary nature of the recorded data. Assuming that an attenuation of the beam by using neutral density filters with factors $C_{i}$, does not change the STID at focus, we can overcome this limitation. Repeating measurements with an additional attenuation by the factor $C_{i}$ of the laser pulse yields a contour that encloses a smaller area because $ I_{th} $ is unaltered. Scaling this contour back to the full intensity of the pulse results in a scaled edge contour intensity:

\begin{equation} \label{eq_contour2}
I_{seci}(x,y,t) = I_{th}\cdot C_{i}.
\end{equation}

\begin{figure}
	\includegraphics[width=\textwidth]{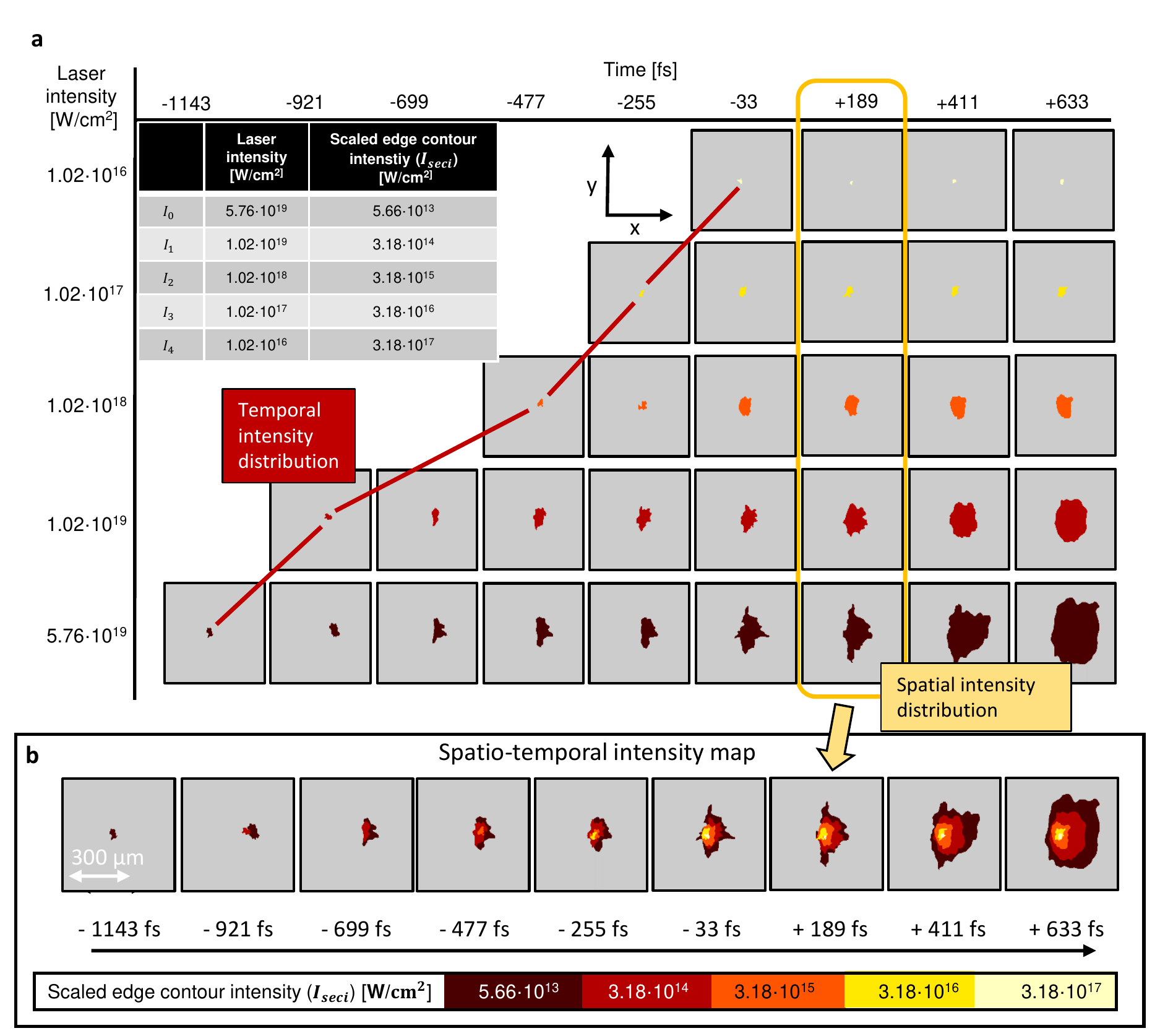}
	\caption{\textbf{Spatio-temporal intensity distribution}. \textbf{a} shows the scaled edge contours exceeding a certain intensity $ I_{seci} $. Each row corresponds to a single-shot with a certain laser intensity (attenuation coefficient) of the laser pulse and thus a different intensity of the scaled edge contour line. While the edges exactly have the depicted intensity, the inner part of the enclosed contour cannot be specified. The table shows the used peak laser intensities and the corresponding measured scaled edge contour intensity. The yellow frame shows the spatial distribution at the peak of the laser pulse $ t=189 $ fs (first frame after t=0) and the red line the temporal intensity distribution of $ x=0$ µm and $ y=15 $ µm. \textbf{b} is the absolute spatio-temporal intensity map and is the main result. Each timestep represents the summation over a column in \textbf{a}.}
	\label{Fig_Spatio-Temporal-Intensity}
\end{figure}

\begin{figure}
	\includegraphics[width=\textwidth]{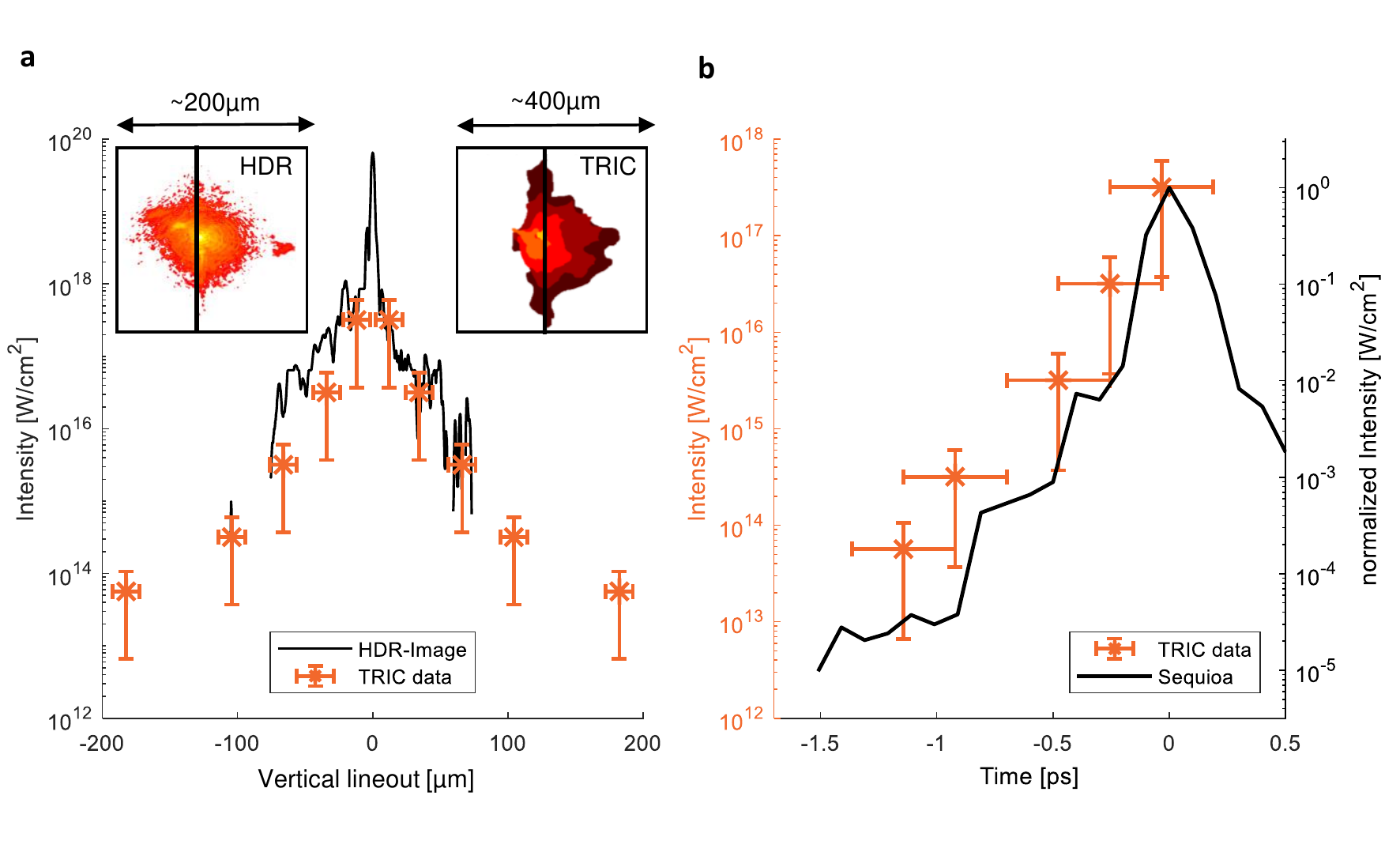}
	\caption{\textbf{Comparison of TRIC to contrast curve and HDR focus image}. \textbf{a}, compares the intensity distribution measured with TRIC and with a HDR camera. Lineouts of the HDR focus (left inset) and the TRIC focus (right inset) are shown. \textbf{b}, shows the temporal contrast of TRIC at $ x=0$ µm and $ y=15 $ µm. For comparison the right axis shows a normalization compared to the measurement taken with the Sequoia-autocorrelator.}
	\label{Fig_Marginals}
\end{figure}

Repeating this for several $C_{i}$ yields the absolute spatio-temporal intensity map as shown in Fig. 3. Each row corresponds to a single-shot with different attenuation coefficient and thus the scaled contour corresponds to a different intensity$ I_{seci} $. The summation over those measurements enables the retrieval of the absolute spatial intensity distribution of the laser focus at distinct time-steps and is shown in Fig. 3b. Fig. 3b is thus the complete STID of a high-power laser pulse in the focal plane, measured at full pulse energy on target and represents our main result. The intensity map features measuring the spatial intensity profiles at select time steps and reveals the temporal dynamics at a given position. The most interesting cases are the spatial distribution of the peak intensity ($ t=0 $) and the temporal contrast at the central point ($ x=0$ and $y=0 $). We therefore evaluated $ I(x,y,t=189$ fs)  as displayed in Fig. \ref{Fig_Marginals}a, since it was the closest timestep past the interaction with the peak of the main pulse ($t=0 $) and $ I(x=0$ µm, $y=15$ µm, t) as displayed in Fig. \ref{Fig_Marginals}b. Due to in its current form limited spatial resolution we could not resolve the central area $ \sim $ 25 µm. Therefore the most intense measured contour represents $3.18\cdot10^{17}$ \si[mode=text]{W.cm^{-2}}, which is orders of magnitude below the classically estimated peak intensity of $5.76\cdot10^{19}$ \si[mode=text]{W.cm^{-2}}.

\subsection{Comparison to contrast curve and laser focus.}
To assess our method we compared the extracted temporal and spatial intensity distribution to data obtained by conventional methods, a contrast profile of a third order auto-correlator (integrated over space) $ g(t) $ and a high-dynamic-range (HDR) image of the laser in the focal plane (integrated over time) $ f(x,y) $\cite{hartmann_spatial_2018}. Both of those standard measurements yield only relative information, which we scale to absolute values by calculating

\begin{equation} \label{eq_Isep}
I(x,y,t) = E_{Laser} \cdot f(x,y) \cdot g(t),
\end{equation}
where $ E_{Laser} = 374$ \si{mJ}  is the full laser pulse energy measured with a power meter.  The spatial shape function of the laser focus $ f(x,y) $ and the temporal laser contrast-curve $ g(t) $ are both normalized, $  \int f(x,y)\,dx\,dy =1$ and $  \int g(t)\,dt =1$. Note that this separation of the temporal and spatial profiles neglects effects of spatio-temporal-coupling\cite{akturk_spatio-temporal_2010} by definition.

\subsection{HDR focus picture.}
The HDR focus image is the integrated spatial distribution over the full duration of the laser pulse and can thus be normalized to $ f(x,y) $. Assuming a Gaussian profile for the temporal distribution of the laser pulse, with a full-width at half-maximum ($\tau_{FWHM}$) of 30 fs (laser pulse duration), the spatial intensity distribution of a measured HDR laser focus at time $ t=0 $ is
\begin{equation} \label{eq_ImaxHdr}
 I(x,y,t=0) =E_{Laser}\cdot f(x,y) \cdot \frac{1}{\sqrt{2\pi}\sigma},
\end{equation}
with $ \sigma=\frac{\tau_{FWHM}}{2.35}  $.

Fig. 4a compares a line-out of this function along the vertical dimension (at x=0) with the result obtained by TRIC at $ t=189 $ fs. The focus, measured offline with the camera, appears continuous since it has more data points, while TRIC contributes a small number of points (dictated by the number of shots with different attenuation coefficients). In the overlap region, both methods yield similar comparable results. With TRIC we can measure down to $ I_{th} $, which exceeds the capability of HDR imaging as this is limited by damage of the camera. With an increased spatial resolution of TRIC, we expect to resolve also the peak of the pulse such that a 7 orders of magnitude dynamic range seems feasible. The horizontal error bar originates from the readout of the diameter of the contour. The rather large error bar for the evaluated intensities is due to the error of the determination in the damage threshold.

\subsection{The temporal intensity distribution.}
The Sequoia is a scanning third order auto-correlator and measures the temporal intensity contrast relative to the laser peak. Comparison to the absolute temporal intensity evolution at a fixed spatial position, as accessible with TRIC, requires more assumptions. We thus solely compare the autocorrelator contrast curve to a normalized curve measured with TRIC as indicated in Fig. \ref{Fig_Marginals}b (right axis). We normalized both intensities such that the peak ($ t=0 $) is equal to one.
The extent of temporal error bars of TRIC is attributed to the time interval correlated to the width of the bandpass filters for each timestep in addition to an increment from time delay uncertainty. The intensity error bars originate from the uncertainty of the determined damage threshold. The TRIC values follow the shape of the autocorrelation trace but indicate reduced temporal contrast. The reason for this is not clear and is likely due to spatio-temporal coupling that only manifests in the focus of the pulse where TRIC is applied. It highlights the necessity of monitoring and controlling the laser contrast on target at full energy, especially in the focal plane. More detailed measurements with an improved spatial resolution will help to shed light on this interesting aspect.

\section*{Summary and Discussion}
By using an elegant and simple method of beam multiplexing we enable recording of nine images of the plasma evolution during the interaction of a high-power laser pulse with a thin solid density target in a single-shot. Our interpretation of the such obtained images in the framework of TRIC yields information about the absolute STID in the focus of a high-power laser pulse. 
 
This is specially interesting in combination with plasma mirrors\cite{obst_efficient_2017} or other nonlinear techniques\cite{hornung_generation_2015} that aim to manipulate the laser pulse close to target. TRIC gives access to the spatial intensity distribution at energies slightly above $ I_{th} $. The measured large diameter here could directly explain the damage extension to the neighboring targets ('fratricide') as often observed in laser-plasma experiments. We note that it requires still analytical effort to account for the horizontal dimension. Because the ignition of the plasma and thus the change of opacity is irreversible on ultra-fast timescales, TRIC only yields information of the rising slope of the laser pulse. Improving the resolution to a few µm will increase the sensitivity to prepulses just above the damage threshold. A more detailed evaluation of the damage threshold would further allow a reduction of the error bar and thus an even more accurate determination of the absolute intensity distribution on target.

The elegance of the nine frame imaging, solely using the natural features of a broadband highly intense laser-pulse, can be adapted to other, more complicated probing techniques, e.g. using holography\cite{chien_single-shot_2000,le_blanc_single-shot_2000} instead of binary shadowgraphy. It can therefore add a temporal component to diverse probing experiments. Thus, we see TRIC as a very first step that is accessible with this technique. 
In the future we foresee use for different pump-probe modalities (e.g. protons, electrons, laser-pulse, x-rays), in particular when those are intrinsically synchronized via laser-driven particle acceleration processes. The demonstrated sub-ps temporal resolution further expands the development of time-of-flight measurement\cite{dromey_picosecond_2016, scuderi_time_2017} of ion bunches with combined temporal and 2D spatial resolution.

\begin{methods}

\subsection{The probe pulse} was coupled out with a pick off mirror at the edge of the main laser pulse in the target chamber (spatial sampling). The timing of the probe beam with respect to the main laser pulse was controlled with a delay stage. The probe beam was then guided through an aperture of 7.5  mm diameter having an energy up to 5 µJ. This aperture was imaged onto the interaction point (IP). A glass rod of 3 cm length was introduced in the beam path prior to target in order to chirp the pulse up to 2 ps. The IP is then further imaged onto a camera via a two lens system.

\subsection{The magnification}
of the optical system is measured with the known aperture in the probe beam. The aperture is imaged onto the IP with a nine times demagnification, resulting in a aperture of 834 µm in the IP. The resulting magnification of the optical probe image was $  6.5 $.

\subsection{The zero timing}
describes the coincidence of the incident of probe and main pulse on the target. It was measured with the use of an air plasma ignited by the attenuated pump pulse in air. A high-power laser pulse can generate an air plasma when a certain intensity threshold \cite{morgan_laser-induced_1975} is reached. Therefore the laser intensity was diminished until the air plasma was solely visible in one or two of the nine frames and therefore marking the peak of the laser pulse with an accuracy of $ \pm 222$ fs .

\subsection{The frame-rate of the camera}
is set by the temporal spacing of the spectral images and the chirp of the probe pulse. Therefore the group velocity dispersion has to be calculated including complete knowledge of the stretching material in the probe beam and respective material characteristics. Since multiple error sources are introduced, we chose a different approach. Two nine-frame images are created, differing in delay by 1000 steps of the motor of the delay stage. If a specific plasma size is observed in two different frames of both nine-frame images, one can check for the change in the frame number and unambiguously correlate frame-number and temporal delay. Knowing that 1000 steps equal a delay of 666 fs the frame-rate can be simply calculated. In the setup described above, a 3 cm glass rod resulted in a delay of 222 fs between each frame and a total observation time of 2 ps within a single-shot.

\subsection{The damage theshold}
was determined by focusing the main laser pulse onto a 200 nm thick plastic foil with a peak intensity that was calculated via equation \ref{eq_ImaxHdr}. The target was examined after the shot with a 20 times magnifying microscope in the vacuum chamber. By reducing the intensity stepwise according to the values given in Fig. 2 we identified that the target was not damaged after a single-shot with peak intensity $ I_{7}=5.77 \cdot 10^{13}$ \si[mode=text]{W.cm^{-2}}, whereas the target showed a clear hole after a shot with $ I_6 $. We thus, supported by the extensions of the damage after a shot with $ I_{6} $, determined the damage threshold of the $ 200$ nm plastic target to be $ I_{th}=6 \pm 5 \cdot 10^{13}$ \si[mode=text]{W.cm^{-2}}. The large error bar could be reduced by more detailed measurements with smaller steps of the intensity variation. We also note, that we determine the damage threshold as indistinguishable from the ionization threshold, breaking the bindings and thus destroying the target. An alteration of the target surface occurs at even lower intensities and is often also referred to as damage.

\end{methods}


\begin{thebibliography}{10}
\expandafter\ifx\csname url\endcsname\relax
  \def\url#1{\texttt{#1}}\fi
\expandafter\ifx\csname urlprefix\endcsname\relax\def\urlprefix{URL }\fi
\providecommand{\bibinfo}[2]{#2}
\providecommand{\eprint}[2][]{\url{#2}}

\bibitem{strickland_compression_1985}
\bibinfo{author}{Strickland, D.} \& \bibinfo{author}{Mourou, G.}
\newblock \bibinfo{title}{Compression of amplified chirped optical pulses}.
\newblock \emph{\bibinfo{journal}{Optics Communications}}
  \textbf{\bibinfo{volume}{56}}, \bibinfo{pages}{219 -- 221}
  (\bibinfo{year}{1985}).

\bibitem{maine_generation_1988}
\bibinfo{author}{Maine, P.}, \bibinfo{author}{Strickland, D.},
  \bibinfo{author}{Bado, P.}, \bibinfo{author}{Pessot, M.} \&
  \bibinfo{author}{Mourou, G.}
\newblock \bibinfo{title}{Generation of ultrahigh peak power pulses by chirped
  pulse amplification}.
\newblock \emph{\bibinfo{journal}{IEEE Journal of Quantum Electronics}}
  \textbf{\bibinfo{volume}{24}}, \bibinfo{pages}{398--403}
  (\bibinfo{year}{1988}).

\bibitem{esarey_physics_2009}
\bibinfo{author}{Esarey, E.}, \bibinfo{author}{Schroeder, C.~B.} \&
  \bibinfo{author}{Leemans, W.~P.}
\newblock \bibinfo{title}{Physics of laser-driven plasma-based electron
  accelerators}.
\newblock \emph{\bibinfo{journal}{Reviews of Modern Physics}}
  \textbf{\bibinfo{volume}{81}}, \bibinfo{pages}{1229--1285}
  (\bibinfo{year}{2009}).

\bibitem{khrennikov_tunable_2015}
\bibinfo{author}{Khrennikov, K.} \emph{et~al.}
\newblock \bibinfo{title}{Tunable {{All}}-{{Optical Quasimonochromatic Thomson
  X}}-{{Ray Source}} in the {{Nonlinear Regime}}}.
\newblock \emph{\bibinfo{journal}{Physical Review Letters}}
  \textbf{\bibinfo{volume}{114}} (\bibinfo{year}{2015}).

\bibitem{roth_bright_2013}
\bibinfo{author}{Roth, M.} \emph{et~al.}
\newblock \bibinfo{title}{Bright {{Laser}}-{{Driven Neutron Source Based}} on
  the {{Relativistic Transparency}} of {{Solids}}}.
\newblock \emph{\bibinfo{journal}{Physical Review Letters}}
  \textbf{\bibinfo{volume}{110}} (\bibinfo{year}{2013}).

\bibitem{kim_radiation_2016}
\bibinfo{author}{Kim, I.~J.} \emph{et~al.}
\newblock \bibinfo{title}{Radiation pressure acceleration of protons to 93
  {{MeV}} with circularly polarized petawatt laser pulses}.
\newblock \emph{\bibinfo{journal}{Physics of Plasmas}}
  \textbf{\bibinfo{volume}{23}}, \bibinfo{pages}{070701}
  (\bibinfo{year}{2016}).

\bibitem{wagner_maximum_2016}
\bibinfo{author}{Wagner, F.} \emph{et~al.}
\newblock \bibinfo{title}{Maximum {{Proton Energy}} above 85 {{MeV}} from the
  {{Relativistic Interaction}} of {{Laser Pulses}} with {{Micrometer Thick CH}}
  2 {{Targets}}}.
\newblock \emph{\bibinfo{journal}{Physical Review Letters}}
  \textbf{\bibinfo{volume}{116}} (\bibinfo{year}{2016}).

\bibitem{snavely_intense_2000}
\bibinfo{author}{Snavely, R.~A.} \emph{et~al.}
\newblock \bibinfo{title}{Intense high-energy proton beams from petawatt-laser
  irradiation of solids}.
\newblock \emph{\bibinfo{journal}{Physical Review Letters}}
  \textbf{\bibinfo{volume}{85}}, \bibinfo{pages}{2945} (\bibinfo{year}{2000}).

\bibitem{daido_review_2012}
\bibinfo{author}{Daido, H.}, \bibinfo{author}{Nishiuchi, M.} \&
  \bibinfo{author}{Pirozhkov, A.~S.}
\newblock \bibinfo{title}{Review of laser-driven ion sources and their
  applications}.
\newblock \emph{\bibinfo{journal}{Reports on Progress in Physics}}
  \textbf{\bibinfo{volume}{75}}, \bibinfo{pages}{056401}
  (\bibinfo{year}{2012}).

\bibitem{macchi_ion_2013}
\bibinfo{author}{Macchi, A.}, \bibinfo{author}{Borghesi, M.} \&
  \bibinfo{author}{Passoni, M.}
\newblock \bibinfo{title}{Ion acceleration by superintense laser-plasma
  interaction}.
\newblock \emph{\bibinfo{journal}{Reviews of Modern Physics}}
  \textbf{\bibinfo{volume}{85}}, \bibinfo{pages}{751--793}
  (\bibinfo{year}{2013}).

\bibitem{schreiber_invited_2016}
\bibinfo{author}{Schreiber, J.}, \bibinfo{author}{Bolton, P.~R.} \&
  \bibinfo{author}{Parodi, K.}
\newblock \bibinfo{title}{Invited {{Review Article}}: ``{{Hands}}-on''
  laser-driven ion acceleration: {{A}} primer for laser-driven source
  development and potential applications}.
\newblock \emph{\bibinfo{journal}{Review of Scientific Instruments}}
  \textbf{\bibinfo{volume}{87}}, \bibinfo{pages}{071101}
  (\bibinfo{year}{2016}).

\bibitem{papadopoulos_apollon_2016}
\bibinfo{author}{Papadopoulos, D.} \emph{et~al.}
\newblock \bibinfo{title}{The {{Apollon}} 10 {{PW}} laser: Experimental and
  theoretical investigation of the temporal characteristics}.
\newblock \emph{\bibinfo{journal}{High Power Laser Science and Engineering}}
  \textbf{\bibinfo{volume}{4}} (\bibinfo{year}{2016}).

\bibitem{_cala_}
\bibinfo{title}{{{CALA}} - {{Centre}} for {{Advanced Laser Applications}}:
  {{Mainpage}}}.
\newblock \bibinfo{howpublished}{https://www.cala-laser.de/}.

\bibitem{higginson_near-100_2018}
\bibinfo{author}{Higginson, A.} \emph{et~al.}
\newblock \bibinfo{title}{Near-100 {{MeV}} protons via a laser-driven
  transparency-enhanced hybrid acceleration scheme}.
\newblock \emph{\bibinfo{journal}{Nature Communications}}
  \textbf{\bibinfo{volume}{9}} (\bibinfo{year}{2018}).

\bibitem{gao_automated_2017}
\bibinfo{author}{Gao, Y.} \emph{et~al.}
\newblock \bibinfo{title}{An automated, 0.5 {{Hz}} nano-foil target positioning
  system for intense laser plasma experiments}.
\newblock \emph{\bibinfo{journal}{High Power Laser Science and Engineering}}
  \textbf{\bibinfo{volume}{5}} (\bibinfo{year}{2017}).

\bibitem{mckenna_high-intensity_2006}
\bibinfo{author}{McKenna, P.} \emph{et~al.}
\newblock \bibinfo{title}{High-intensity laser-driven proton acceleration:
  Influence of pulse contrast}.
\newblock \emph{\bibinfo{journal}{Philosophical Transactions of the Royal
  Society A: Mathematical, Physical and Engineering Sciences}}
  \textbf{\bibinfo{volume}{364}}, \bibinfo{pages}{711--723}
  (\bibinfo{year}{2006}).

\bibitem{blount_recovery_1969}
\bibinfo{author}{Blount, E.~I.} \& \bibinfo{author}{Klauder, J.~R.}
\newblock \bibinfo{title}{Recovery of {{Laser Intensity}} from {{Correlation
  Data}}}.
\newblock \emph{\bibinfo{journal}{Journal of Applied Physics}}
  \textbf{\bibinfo{volume}{40}}, \bibinfo{pages}{2874--2875}
  (\bibinfo{year}{1969}).

\bibitem{luan_high_1993}
\bibinfo{author}{Luan, S.}, \bibinfo{author}{Hutchinson, M. H.~R.},
  \bibinfo{author}{Smith, R.~A.} \& \bibinfo{author}{Zhou, F.}
\newblock \bibinfo{title}{High dynamic range third-order correlation
  measurement of picosecond laser pulse shapes}.
\newblock \emph{\bibinfo{journal}{Measurement Science and Technology}}
  \textbf{\bibinfo{volume}{4}}, \bibinfo{pages}{1426--1429}
  (\bibinfo{year}{1993}).

\bibitem{wang_single-shot_2015}
\bibinfo{author}{Wang, Y.} \emph{et~al.}
\newblock \bibinfo{title}{Single-shot measurement of $>$1010 pulse contrast for
  ultra-high peak-power lasers}.
\newblock \emph{\bibinfo{journal}{Scientific Reports}}
  \textbf{\bibinfo{volume}{4}} (\bibinfo{year}{2015}).

\bibitem{oksenhendler_high_2017}
\bibinfo{author}{Oksenhendler, T.}, \bibinfo{author}{Bizouard, P.},
  \bibinfo{author}{Albert, O.}, \bibinfo{author}{Bock, S.} \&
  \bibinfo{author}{Schramm, U.}
\newblock \bibinfo{title}{High dynamic, high resolution and wide range single
  shot temporal pulse contrast measurement}.
\newblock \emph{\bibinfo{journal}{Optics Express}}
  \textbf{\bibinfo{volume}{25}}, \bibinfo{pages}{12588} (\bibinfo{year}{2017}).

\bibitem{hartmann_spatial_2018}
\bibinfo{author}{Hartmann, J.} \emph{et~al.}
\newblock \bibinfo{title}{The spatial contrast challenge for intense
  laser-plasma experiments}.
\newblock In \emph{\bibinfo{booktitle}{The Proceedings of the 6th {{Target
  Fabrication Workshop}} ({{TFW6}}) and the {{Targetry}} for {{High Repetition
  Rate Laser}}-{{Driven Sources}} ({{Targ3}}) {{Conference}} ({{Accepted}})}}
  (\bibinfo{address}{Salamanca}, \bibinfo{year}{2018}).

\bibitem{pariente_spacetime_2016}
\bibinfo{author}{Pariente, G.}, \bibinfo{author}{Gallet, V.},
  \bibinfo{author}{Borot, A.}, \bibinfo{author}{Gobert, O.} \&
  \bibinfo{author}{Qu\'er\'e, F.}
\newblock \bibinfo{title}{Space\textendash{}time characterization of
  ultra-intense femtosecond laser beams}.
\newblock \emph{\bibinfo{journal}{Nature Photonics}}
  \textbf{\bibinfo{volume}{10}}, \bibinfo{pages}{547--553}
  (\bibinfo{year}{2016}).

\bibitem{seuferling_efficient_2017}
\bibinfo{author}{Seuferling, S.} \emph{et~al.}
\newblock \bibinfo{title}{Efficient offline production of freestanding thin
  plastic foils for laser-driven ion sources}.
\newblock \emph{\bibinfo{journal}{High Power Laser Science and Engineering}}
  \textbf{\bibinfo{volume}{5}} (\bibinfo{year}{2017}).

\bibitem{gabolde_single-frame_2008}
\bibinfo{author}{Gabolde, P.} \& \bibinfo{author}{Trebino, R.}
\newblock \bibinfo{title}{Single-frame measurement of the complete
  spatiotemporal intensity and phase of ultrashort laser pulses using
  wavelength-multiplexed digital holography}.
\newblock \emph{\bibinfo{journal}{Journal of the Optical Society of America B}}
  \textbf{\bibinfo{volume}{25}}, \bibinfo{pages}{A25} (\bibinfo{year}{2008}).

\bibitem{nakagawa_sequentially_2014}
\bibinfo{author}{Nakagawa, K.} \emph{et~al.}
\newblock \bibinfo{title}{Sequentially timed all-optical mapping photography
  ({{STAMP}})}.
\newblock \emph{\bibinfo{journal}{Nature Photonics}}
  \textbf{\bibinfo{volume}{8}}, \bibinfo{pages}{695--700}
  (\bibinfo{year}{2014}).

\bibitem{siders_efficient_1998}
\bibinfo{author}{Siders, C.~W.}, \bibinfo{author}{Siders, J. L.~W.},
  \bibinfo{author}{Taylor, A.~J.}, \bibinfo{author}{Park, S.-G.} \&
  \bibinfo{author}{Weiner, A.~M.}
\newblock \bibinfo{title}{Efficient high-energy pulse-train generation using a
  2\^n-pulse michelson interferometer}.
\newblock \emph{\bibinfo{journal}{Applied Optics}}
  \textbf{\bibinfo{volume}{37}}, \bibinfo{pages}{5302} (\bibinfo{year}{1998}).

\bibitem{kaluza_time-sequence_2008}
\bibinfo{author}{Kaluza, M.~C.}, \bibinfo{author}{Santala, M. I.~K.},
  \bibinfo{author}{Schreiber, J.}, \bibinfo{author}{Tsakiris, G.~D.} \&
  \bibinfo{author}{Witte, K.~J.}
\newblock \bibinfo{title}{Time-sequence imaging of relativistic
  laser\textendash{}plasma interactions using a novel two-color probe pulse}.
\newblock \emph{\bibinfo{journal}{Applied Physics B}}
  \textbf{\bibinfo{volume}{92}}, \bibinfo{pages}{475--479}
  (\bibinfo{year}{2008}).

\bibitem{green_single_2014}
\bibinfo{author}{Green, J.~S.} \emph{et~al.}
\newblock \bibinfo{title}{Single shot, temporally and spatially resolved
  measurements of fast electron dynamics using a chirped optical probe}.
\newblock \emph{\bibinfo{journal}{Journal of Instrumentation}}
  \textbf{\bibinfo{volume}{9}}, \bibinfo{pages}{P03003--P03003}
  (\bibinfo{year}{2014}).

\bibitem{soong_experimental_2011}
\bibinfo{author}{Soong, K.}, \bibinfo{author}{Colby, E.~R.} \&
  \bibinfo{author}{McGuinness, C.}
\newblock \bibinfo{title}{Experimental {{Determination}} of {{Damage Threshold
  Characteristics}} of {{IR Compatible Optical Materials}}}.
\newblock \emph{\bibinfo{journal}{New York}} \bibinfo{pages}{3}
  (\bibinfo{year}{2011}).

\bibitem{akturk_spatio-temporal_2010}
\bibinfo{author}{Akturk, S.}, \bibinfo{author}{Gu, X.},
  \bibinfo{author}{Bowlan, P.} \& \bibinfo{author}{Trebino, R.}
\newblock \bibinfo{title}{Spatio-temporal couplings in ultrashort laser
  pulses}.
\newblock \emph{\bibinfo{journal}{Journal of Optics}}
  \textbf{\bibinfo{volume}{12}}, \bibinfo{pages}{093001}
  (\bibinfo{year}{2010}).

\bibitem{obst_efficient_2017}
\bibinfo{author}{Obst, L.} \emph{et~al.}
\newblock \bibinfo{title}{Efficient laser-driven proton acceleration from
  cylindrical and planar cryogenic hydrogen jets}.
\newblock \emph{\bibinfo{journal}{Scientific Reports}}
  \textbf{\bibinfo{volume}{7}} (\bibinfo{year}{2017}).

\bibitem{hornung_generation_2015}
\bibinfo{author}{Hornung, M.} \emph{et~al.}
\newblock \bibinfo{title}{Generation of 25-{{TW Femtosecond Laser Pulses}} at
  515 nm with {{Extremely High Temporal Contrast}}}.
\newblock \emph{\bibinfo{journal}{Applied Sciences}}
  \textbf{\bibinfo{volume}{5}}, \bibinfo{pages}{1970--1979}
  (\bibinfo{year}{2015}).

\bibitem{chien_single-shot_2000}
\bibinfo{author}{Chien, C.~Y.} \emph{et~al.}
\newblock \bibinfo{title}{Single-shot chirped-pulse spectral interferometry
  used to measure the femtosecond ionization dynamics of air}.
\newblock \emph{\bibinfo{journal}{Optics Letters}}
  \textbf{\bibinfo{volume}{25}}, \bibinfo{pages}{578} (\bibinfo{year}{2000}).

\bibitem{le_blanc_single-shot_2000}
\bibinfo{author}{Le~Blanc, S.~P.}, \bibinfo{author}{Gaul, E.~W.},
  \bibinfo{author}{Matlis, N.~H.}, \bibinfo{author}{Rundquist, A.} \&
  \bibinfo{author}{Downer, M.~C.}
\newblock \bibinfo{title}{Single-shot measurement of temporal phase shifts by
  frequency-domain holography}.
\newblock \emph{\bibinfo{journal}{Optics Letters}}
  \textbf{\bibinfo{volume}{25}}, \bibinfo{pages}{764} (\bibinfo{year}{2000}).

\bibitem{dromey_picosecond_2016}
\bibinfo{author}{Dromey, B.} \emph{et~al.}
\newblock \bibinfo{title}{Picosecond metrology of laser-driven proton bursts}.
\newblock \emph{\bibinfo{journal}{Nature Communications}}
  \textbf{\bibinfo{volume}{7}}, \bibinfo{pages}{10642} (\bibinfo{year}{2016}).

\bibitem{scuderi_time_2017}
\bibinfo{author}{Scuderi, V.} \emph{et~al.}
\newblock \bibinfo{title}{Time of {{Flight}} based diagnostics for high energy
  laser driven ion beams}.
\newblock \emph{\bibinfo{journal}{Journal of Instrumentation}}
  \textbf{\bibinfo{volume}{12}}, \bibinfo{pages}{C03086--C03086}
  (\bibinfo{year}{2017}).

\bibitem{morgan_laser-induced_1975}
\bibinfo{author}{Morgan, C.~G.}
\newblock \bibinfo{title}{Laser-induced breakdown of gases}.
\newblock \emph{\bibinfo{journal}{Reports on Progress in Physics}}
  \textbf{\bibinfo{volume}{38}}, \bibinfo{pages}{621--665}
  (\bibinfo{year}{1975}).

\end{thebibliography}


\begin{addendum}
 \item The DFG-funded Cluster of Excellence Munich-Centre for Advanced Photonics (MAP) supported this work. We want to thank Prof. Stefan Karsch and his great team, Max Gilljohann, Hao Ding, Johannes Goetzfried, Gregor Schilling, Sabine Schindler for there tremendous effort in building, operating and improving the ATLAS300 laser. We also want to thank Peter Hilz, Matthias Otto Haug, Daniel Hahner, Sebastian Seuferling, Jerzy Szerypo and Simon Storck for excellent support concerning the plasma targets. We also want to thank Paul Bolton for excellent support composing the manuscript.
 \item[Author Contribution] T. M. O., C. K. and D. H. designed and built the experimental infrastructure in LEX-Photonics. D. H., M. S. and J. H. B. designed and built the experimental infrastructure. K. A. and J. S. developed the idea and first tests for the multi-frame imaging technique. D. H., M. S., J. H. B., J. H. and E. R. performed the experiments. D. H., M. S., J. H. B. and J. S analyzed the data, discussed and interpreted the results and prepared the manuscript. 
 \item[Competing Interests] I declare that the authors have no competing interests,or other interests that might be perceived to influence the results and/or discussion reported in this paper.
 \item[Correspondence] Correspondence and requests for materials
should be addressed to Daniel Haffa (email: Daniel.Haffa@physik.lmu.de, Jianhui Bin (jianhuibin@lbl.gov) or Martin Speicher(Martin.Speicher@physik.lmu.de)
\end{addendum}

\end{document}